\documentclass{article}

\usepackage{authblk}
\usepackage[utf8]{inputenc}
\usepackage[left=2cm,right=2cm,top=1.5cm,bottom=1.5cm]{geometry}
\usepackage{amsmath}
\usepackage{amssymb}
\usepackage{graphicx}
\usepackage{subfig}
\usepackage{mathtools}
\usepackage{mathpazo}
\usepackage[mathpazo]{flexisym}
\usepackage{breqn}
\usepackage{verbatim}
\usepackage{xcolor}
\usepackage{ulem}
\usepackage{placeins}
\usepackage{hyperref}
\usepackage{soul}
\hypersetup{
    colorlinks=true,
    linkcolor=blue,
    filecolor=magenta,      
    urlcolor=cyan,
    citecolor=teal
    }
\usepackage{comment}
\graphicspath{{figs/}}

\usepackage[style=phys,url=true]{biblatex}
\DeclareMathOperator{\sech}{sech}

\title{Kink Dynamics in a Non-Autonomous Sine-Gordon Model}
\author[1]{Tomasz Dobrowolski}
\author[2]{Jacek Gatlik}
\author[2]{Zofia Bryłowska}
\author[3]{Panayotis G. Kevrekidis}
\affil[1]{\textit{Department of Physics and Applied Mathematics, University of the National Education Commission in Krakow, Podchor\c{a}\.zych  2, 30-084 Cracow, Poland}}
\affil[2]{\textit{AGH University of Krakow, Faculty of Physics and Applied Computer Science, 30-059 Krakow, Poland}}
\affil[3]{\textit{Department of Mathematics and Statistics, University of Massachusetts, Amherst, Massachusetts 01003-4515, USA}}
\date{\today}

\begin{document}
\maketitle

\begin{abstract}
The sine-Gordon model with space- and time-dependent parameters is considered. A highly
accurate effective model with two degrees of freedom is constructed, allowing the description of the kink movement in this model even for extremely long times and nontrivial trajectories of the coherent structure. As a stringent test of the reduced order
model, the case of a temporal drive leading
to extremely complex kink motion is studied. 
The two-degree-of-freedom approximation is found to
faithfully reproduce the behavior of the full field-theoretic
model paving the way for both deeper understanding and improved design of soliton-based devices. 

\end{abstract} \hspace{10pt}

{\bf The sine-Gordon model is a prototypical nonlinear classical field theory at the center of the analysis of solitary wave dynamics, especially in describing the interactions of the solitary waves. It arises in a wide range of applications
from systems as simple as coupled torsion pendula to ones
as complex
as superconducting Josephson junctions.
It has also been a focal point, notably in its
discrete so-called Frenkel-Kontorova form, 
of one of the many significant contributions
of S. Aubry's illustrious career. 
Solitons frequently emerge in nonlinear dynamical systems as robust, localized excitations. In many physical contexts these localized structures play a fundamental role in determining the dynamical and transport properties of the system. In most macroscopic systems, however, the presence of dissipation and external driving forces calls for the consideration of perturbed variants of the sine-Gordon model to accurately capture the underlying dynamics. To gain insight into the fundamental features of such complex field theories, it is often advantageous to reduce their infinite degrees of freedom to a handful of collective coordinates that effectively capture the core dynamics.
These reduced models offer a tractable framework to describe kink dynamics and, by extension, to analyze and even design more efficient devices governed by the corresponding field-theoretic
model. Our aim herein is to exploit the connection between the full sine-Gordon model and its low-dimensional reductions, proposing not only an effective two-degree-of-freedom description, but, crucially, a model that faithfully reproduces the dynamics of the full field-theoretic system. }

\section{Introduction}

Solitons are highly robust localized wave structures that preserve their shape while propagating at a constant velocity, emerging from a delicate balance between nonlinear and dispersive effects in the medium~\cite{DauxoisPeyrard10,ablowitz2011nonlinear}. Initially observed in shallow water waves~\cite{Ablowitz1991,ablowitz2011nonlinear}, solitons have since been identified in a wide range of physical, chemical, and biological systems, providing profound insights into natural phenomena and advancing various technological applications.
For example, in magnetic materials, the sine-Gordon model is instrumental in understanding phenomena such as the Kosterlitz-Thouless transition in two-dimensional systems, framed within the continuous classical XY model \cite{Kosterlitz1973}. Similarly, in Josephson junctions, the sine-Gordon equation governs the dynamics of gauge-invariant phase differences, elucidating the behavior of long Josephson junctions and their significance in quantum computing and superconducting electronics \cite{Lomdahl1985}.
The sine-Gordon model has also been a central theme of
focus for the work of S. Aubry~\cite{Aubry1983}.
Most notably, in its well-known
discrete form (as the Frenkel-Kontorova model,
relevant to dislocation theory~\cite{Frenkel1939})
Aubry and collaborators~\cite{Aubry1983,MPeyrard_1983}
established the transition by breaking of 
analyticity for its ground states. 

Extensions of soliton-based applications have also arisen
in polymer physics, where solitons contribute to energy transfer along polymer chains, offering a microscopic perspective on structural defects, their mobility, and associated transitions \cite{Heeger1988}. This understanding is crucial for engineering materials with tailored mechanical and electrical properties.  
Furthermore, in biological systems, solitons have been argued to play a fundamental role in low-frequency collective motions within proteins and DNA, facilitating energy transfer and conformational changes essential for enzymatic activity and genetic regulation \cite{Davydov1977,Davydov1985,peyrard1989statistical}. Their presence underscores the importance of nonlinear excitations in biological functionality.
From a more practical/technological perspective, in fiber optics, solitons balance dispersion and nonlinearity, enabling long-distance data transmission without signal degradation, forming the basis of high-capacity optical communication networks \cite{Agrawal2019}. Cavity solitons in photonic devices have been explored for optical memory and information processing, allowing for reconfigurable optical memory arrays where individual solitons can be written, erased, and manipulated for all-optical buffering and computing \cite{Ankiewicz2008}.
Frequently described by the sine-Gordon model, e.g., as concerns superconducting Josephson junctions, coupled torsion pendula, or surfaces of constant negative curvature, among others \cite{Barone1971,CKW14}, solitons remain at the forefront of scientific and technological exploration across diverse scientific domains.

In recent years, there has been a growing interest, e.g., in condensed-matter settings, 
towards exploring the far-from-equilibrium evolution of physical systems. This interest is largely motivated by progress in cold-atom experiments. By applying different methods, the system can be pushed out of equilibrium, posing the fundamental challenge of understanding its long-term behavior.  In particular, parametric resonance in the atomic Bose-Einstein condensate has recently been the subject of intensive research. This research is both theoretical \cite{Nicolin2007,Ripoll1999,Tozzo2005,Kramer2005} and experimental \cite{Modugno2006,Schori2004,engels,choi} and has, by now, spanned multiple
decades. The possibility of experimental exploration of this system is a consequence of the relatively straightforward generation of periodic disturbances, for example, through periodic changes in the parameters of the magnetic trap (as well as in those of
the nonlinearity via the so-called Feshbach resonances~\cite{chin}).

In the present work we investigate a modified sine-Gordon model that contains two coordinate-dependent functions. 
One was introduced to describe the curvature of the  Josephson junction \cite{Gatlik2021}, and the other to describe changes in the thickness of the junction 
\cite{McLaughlin1978}. 
More specifically here, we study a more general model, in the sense that it allows the parameter preceding the  potential term to be explicitly time-dependent. 
Our goal is to obtain an effective description based on only two degrees of freedom that would reliably describe the behavior of a field system that is significantly modified in relation to the complete sine-Gordon model. Naturally, we are interested in the description in the sector where kink solutions are present.
In what follows, we will also consider on the impact of periodic forcing on the dynamics of the kink. To be precise, the disturbance we consider has the form of a wave running at a fixed wave vector and frequency. This type of disturbance introduces highly non-trivial behavior into the system.

The article is organized as follows. In section II, we define the model to be examined. In this part, we determine the form of the functions that modify the sine-Gordon model as well as the initial and boundary conditions that are used in numerical simulations.
In the third part (section III), we focus on the case where there is no dissipation in the system (and no external forcing). In this case, we obtain an effective description, which is tested in various situations. The investigations cover both the case where there is no explicit time dependence and the case of the potentially quite complex evolution where the parameter preceding the potential term contains a periodic time dependence. Finally, we take into consideration the presence of dissipation in the system. 

\section{System description}
The most general form of the sine-Gordon model that we consider herein is
a partial differential equation (PDE) of the form:
\begin{equation}
\label{sine-gordon+}
\partial_t^2\phi + \eta \partial_t \phi - \partial_x(\mathcal{F}(x)\partial_x\phi) + g(t,x) \sin(\phi)  = - \Gamma,
\end{equation}
where
\begin{equation}
\label{f_function}
\mathcal{F}(x) = 1 + \varepsilon_2\sin\left(\frac{ \pi }{12} x\right) \mathrm{ ~~~~ and ~~~~ } g(t,x) =1+\varepsilon_1\sin(k x -\omega t).
\end{equation}
The constant $\eta$ in the equation describes the dissipation present in the system. {TD In a Josephson junction, it is related to the existence of quasiparticle current. Within this framework, the external forcing $\Gamma$
corresponds to the bias current applied to the junction. The function $\mathcal{F}(x)$ was introduced in \cite{Dobrowolski2012,Dobrowolski2009,Gatlik2021} to account for the effects associated with the curvature of the junction. In the context of Josephson junctions, the modulation function $g(t,x)$ represents the presence of an
external alternating electromagnetic field, which periodically enhances or suppresses the Josephson response by modulating the amplitude of the critical current.  The experimental study of the impact of microwave radiation on the Josephson junctions was undertaken in \cite{Barkov2004}. The article reports the motion of Josephson vortices induced by microwave radiation. Experimental confirmation of this effect was achieved using low-temperature laser scanning microscopy. Moreover, the analysis shows strongly irregular vortex motion. In the present work, we propose a mechanism by which variations in the critical current can induce fluxon motion.}

Note that $g$ is not only a function of space but also depends on time. In fact, the time dependence of the function $g$ has the form of a wave running at a wave vector $k$ and a frequency $\omega$. 
A perturbation of this type in certain parameter ranges allows the transport of kink between minima of the spatial potential without the presence of an external forcing, {i.e.}, for $\Gamma=0$. 
We take the initial conditions in this system as
\begin{equation}\label{phi_wp1}
\phi(0,x)=4 \arctan \left[ \exp \left( \sqrt{\frac{g(t=0,x_0)}{{\mathcal{F}(x_0)}}} \gamma_0 \, (x-x_0) \right) \right],
\end{equation}
\begin{equation}\label{phi_wp2}
\partial_t \phi(0,x)= - 2 v\gamma_0 \sqrt{ \frac{g(t=0,x_0)}{\mathcal{F}(x_0)}}\sech \left[  \sqrt{\frac{g(t=0,x_0)}{\mathcal{F}(x_0)}} \gamma_0 \, (x-x_0) \right],
\end{equation}
where $\gamma_0=1/\sqrt{1-v^2}$. 
This choice of initial conditions allows to harmonize the considered field configuration with the background formed by the various inhomogeneities present in the system and the ansatz assumed in the effective description.
This choice of initial conditions minimizes the amount of energy stored in the field configuration. It allows us to avoid both excitation and the accompanying radiation of energy from the field configuration.
Additionally we assume standard, and consistent with the initial
conditions  kinklike Dirichlet boundary conditions.

\section{Non Dissipative Case}

\subsection{Effective description}
For the dissipationless case and assuming that
the forcing $\Gamma$ is zero, the field equation of the model \eqref{sine-gordon+} reduces to the form
\begin{equation}
\label{sine-gordon}
\partial_t^2 \phi - \partial_x (\mathcal{F}(x)\partial_x \phi) + g(t,x) \sin \phi = 0.
\end{equation}
The construction of the effective description for this system is based on the Lagrangian
\begin{equation}\label{L}
L= \int_{-\infty}^{+\infty} dx {\cal L}(\phi)
=\int_{-\infty}^{+\infty} dx  \left[ \frac{1}{2}~ (\partial_t
\phi)^2 - \frac{1}{2} ~{\cal F}(x) (\partial_x \phi)^2 - g(t,x)
(1-\cos \phi) \right].
\end{equation}
In order to obtain an effective model describing the dynamics of the kink, we switch from the field variable $\phi(t,x)$ to the field variable $\xi(t,x)$ though transformation
\begin{equation}\label{phi}
\phi(t,x)=4 \arctan e^{\xi(t,x)}.
\end{equation}
The above Lagrangian written with the new field variable $\xi$ reduces to the form
\begin{equation}\label{Lxi}
L=4 \int_{-\infty}^{+\infty} dx  \,\, {\rm sech}^2 \xi \left[
\frac{1}{2}\,(\partial_{t} \xi)^2 - \frac{1}{2}\, {\cal
F}(x)(\partial_{x} \xi)^2 -\frac{1}{2} ~g(t,x) \right].
\end{equation}
In the last formula we have used the identity of $1-\cos \phi = 2 {\rm sech}^2 \xi$  satisfied for the function described by
Eq.~\eqref{phi}.
Restricting the dynamics to the sector containing the kink solution is realized when the ansatz is assumed
\begin{equation}\label{xi}
 \xi(t,x)= \sqrt{ \frac{g(t,x_0(t))}{{\cal
F}(x_0(t))}} \,\, \gamma(t) \, (x-x_0(t)).
\end{equation}
Naturally, at this point, we reduce the dynamics of a field system with an infinite number of degrees of freedom to a system that is described with only two degrees of freedom. The first degree of freedom  
$x_0(t)$ describes the position of the center of the kink understood as the place where the function $\phi$ takes the value $\pi$. The second degree of freedom $\gamma(t)$ describes the inverse of the thickness (i.e., width)
of the kink.
The second variable describes any effects, both dynamic and kinematic, affecting the thickness of the kink.
Performing the integration over the spatial variable leads to an expression for the effective Lagrangian: 
\begin{equation}\label{Leff}
L_{eff} = \frac{1}{2}\, M \Dot{x}_0^2 +\frac{1}{2}\, m
\Dot{\gamma}^2 - \kappa \Dot{x}_0 \Dot{\gamma} + \alpha \Dot{x}_0
- \beta \Dot{\gamma} - V.
\end{equation}
The parameters present in the Lagrangian are defined by the corresponding integrals in Appendix.
Naturally, all coefficients appearing in the Lagrangian are functions of the variables $x_0$ and $\gamma$. In addition, due to the explicit dependence of the function g on time, all coefficients are also explicitly time-dependent.
The equations of motion obtained for this Lagrangian take the form:
\begin{equation}
\begin{gathered}
\label{2dof_ansatz}
    M\Ddot{x}_0-\kappa\Ddot{\gamma}+\frac{1}{2}(\partial_{x_0}M)\Dot{x}_0^2-\frac{1}{2}(\partial_{x_0}m)\Dot{\gamma}^2-(\partial_{\gamma}\kappa)\Dot{\gamma}^2+(\partial_{\gamma}M)\Dot{\gamma}\Dot{x}_0+ \\
    (\partial_t M) \Dot{x}_0 - (\partial_t \kappa) \Dot{\gamma} + (\partial_{\gamma} \alpha) \Dot{\gamma} + (\partial_{x_0} \beta) \Dot{\gamma} + \partial_t \alpha +   \partial_{x_0}V=0,\\
    m\Ddot{\gamma}-\kappa\Ddot{x}_0 +\frac{1}{2}(\partial_{\gamma}m)\Dot{\gamma}^2-\frac{1}{2}(\partial_{\gamma}M)\Dot{x}_0^2-(\partial_{x_0}\kappa)\Dot{x}_0^2+(\partial_{x_0}m)\Dot{x}_0\Dot{\gamma}+\\
(\partial_t m) \Dot{\gamma} - (\partial_t \kappa) \Dot{x}_0 - (\partial_{x_0} \beta) \Dot{x}_0 - (\partial_{\gamma} \alpha) \Dot{x_0} - \partial_t \beta +   \partial_{\gamma}V=0.
\end{gathered}
\end{equation}
Obviously, the partial derivatives with respect to time appearing in this equation are only related to the explicit time dependence entering in all coefficients through the g function.

\subsection{Numerical results in the absence of an explicit time dependence }

\subsubsection{The case of $\varepsilon_2 = 0$ and $\omega = 0$}
\begin{figure}[h!]
    \centering
    \subfloat{{\includegraphics[height=4.5cm]{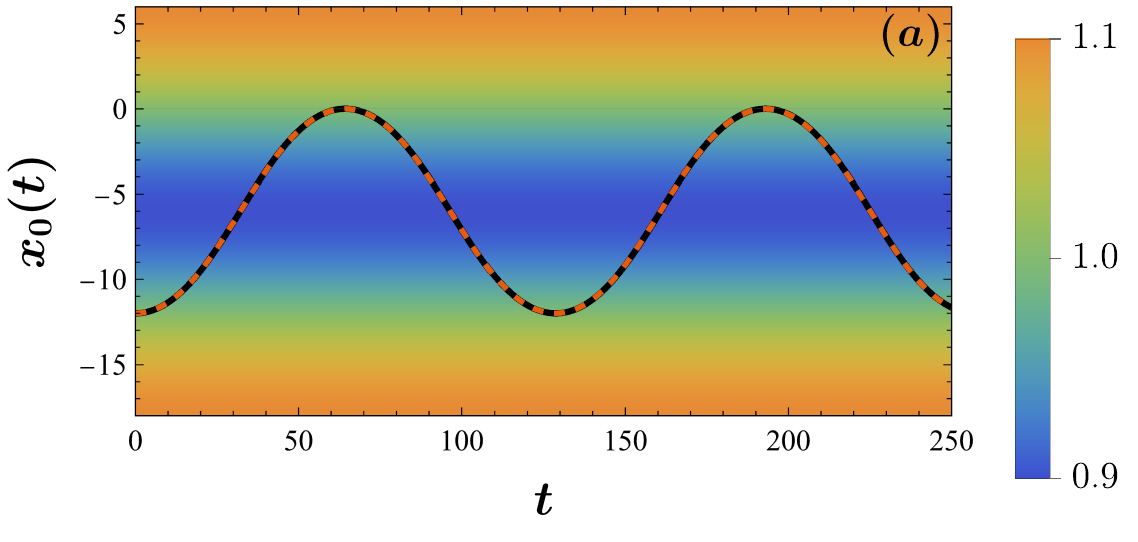}}}
    \quad
    \subfloat{{\includegraphics[height=4.5cm]{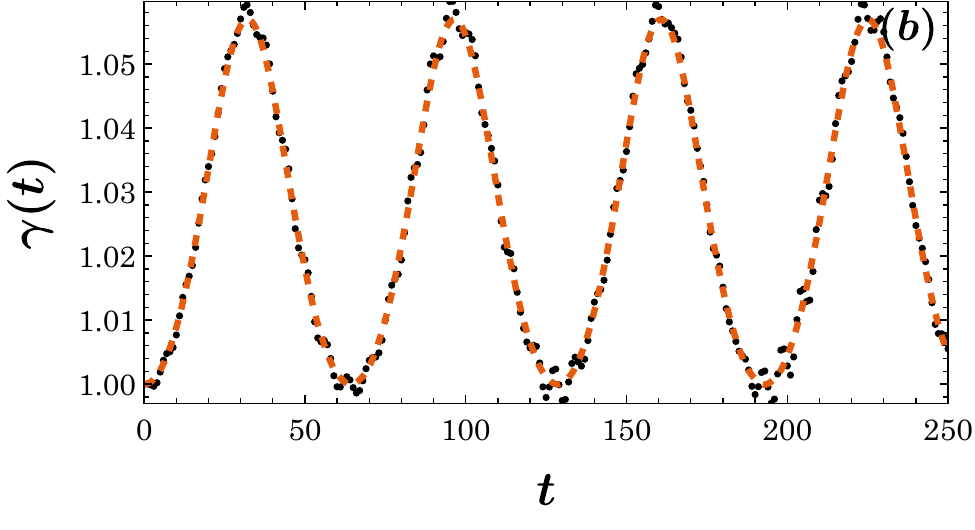}}}
    \quad
    \subfloat{{\includegraphics[height=4.5cm]{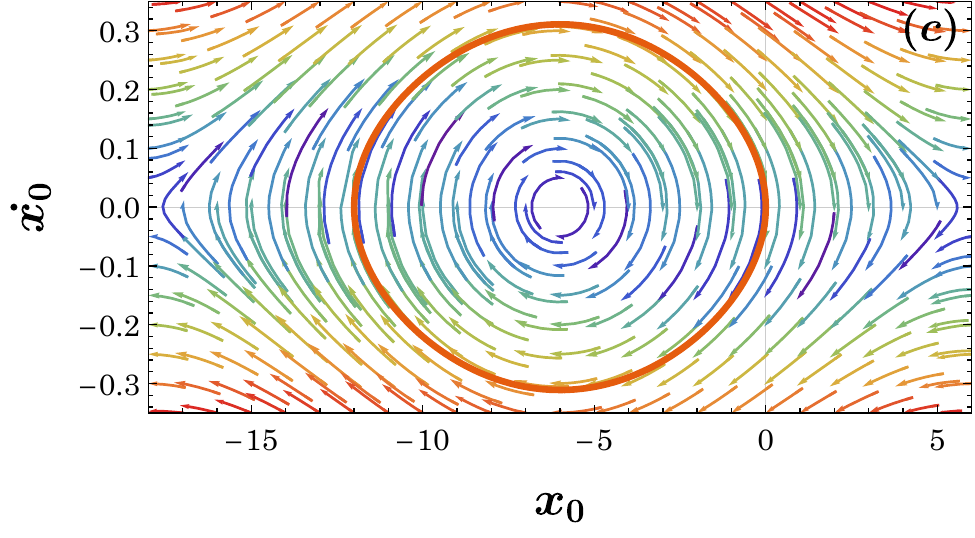}}}
    \caption{Figure (a): A comparison of the kink's position for the solution obtained from the original field-theoretic model \eqref{sine-gordon} depicted by black line and the model described by equations \eqref{2dof_ansatz}, represented by orange dashed line, in the periodic potential. The colors in this figure represent the values of the function $g(x)$ (as shown in the legend to the right). Figure (b): The evolution of the variable $\gamma(t)$ obtained from the PDE model (black points) and the effective model (orange dashed line). Figure (c): Phase portrait of the system in the cross-section $\gamma=1$ and $\Dot{\gamma}=0$. The orange line illustrates the trajectory displayed in the top figure. Here, $\varepsilon_1 = 0.1$, $\varepsilon_2 = 0$, $k=\pi/12$, the initial velocity is $v=0$, and $x_0=-12$.}
    \label{fig_01}
\end{figure}

To begin, we consider the case in which deformations in the gradient term are absent, while spatially periodic changes in the potential term in the Lagrangian \eqref{L} are present, i.e., ${\cal F}(x)=1$ and $g(x)=1+\varepsilon_2 \sin{k x}$. 
In the first simulation, the kink initially rests at the position $x_0=-12$. This location relative to the waveform of the function $g$ is shown in Figure \ref{fig_01} (a). The legend on the right of the diagram refers to the values of the $g$ function. This figure shows the kink trajectory obtained from the field model \eqref{sine-gordon} (black line) and on the basis of the effective model \eqref{2dof_ansatz} (orange dashed line). 
As can be seen, the reduced system of ordinary differential equations
(ODEs) captures very accurately the dynamics
of the original field-theoretic model. We have a similar situation in the case of panel (b), where the time dependence of the $\gamma$ variable is shown. The black points in this figure were obtained from the field model while the orange dashed line from the effective model. One can see the time shift of the $x_0=x_0(t)$ dependence compared to the $\gamma=\gamma(t)$ function shown in panel (b). It can be observed that at instants when $x_0$ reaches the turning points (i.e., the kink stops) $\gamma$ becomes equal to 1. On the other hand, $\gamma$ reaches its maximum when the kink has the highest velocity passing through a point equidistant from both turning points. Obviously, at this point we are dealing with the greatest relativistic contraction.
Indeed, in the described process, the gamma variable 
can be thought of as capturing
the kinematic effects (analogously to the Lorentz factor).
Panel (c) of this figure  shows the phase portrait of the system in the cross section $\gamma = 1$ and $\Dot{\gamma} = 0$. The orange line illustrates the trajectory shown in Figure \ref{fig_01} (a). In this Figure  we assumed $\varepsilon_1=0.1$, $\varepsilon_2=0$, and $k=\pi/6$.

\begin{figure}[h!]
    \centering
    \subfloat{{\includegraphics[height=4.5cm]{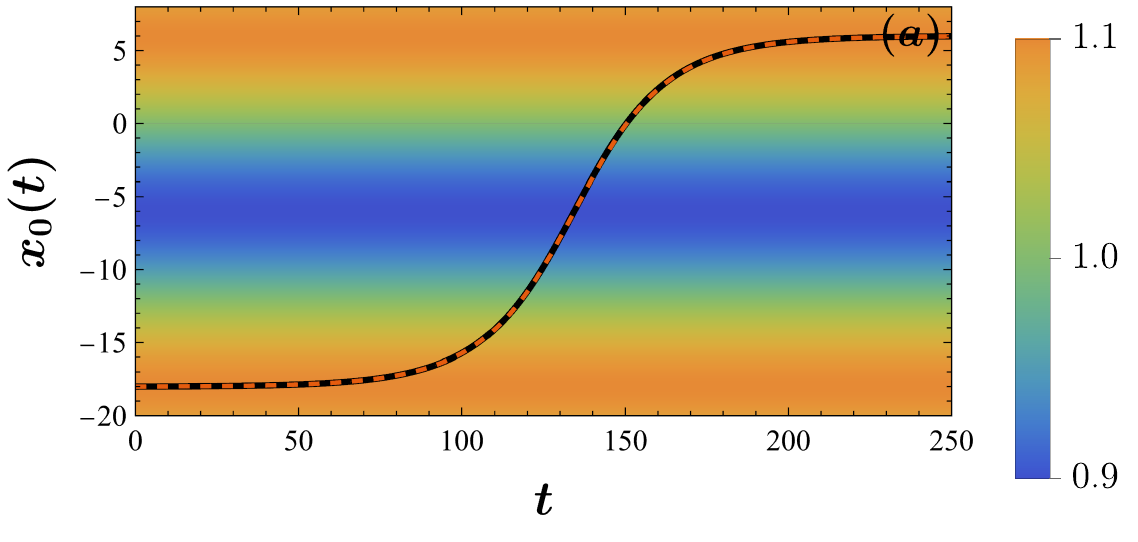}}}
    \quad
    \subfloat{{\includegraphics[height=4.5cm]{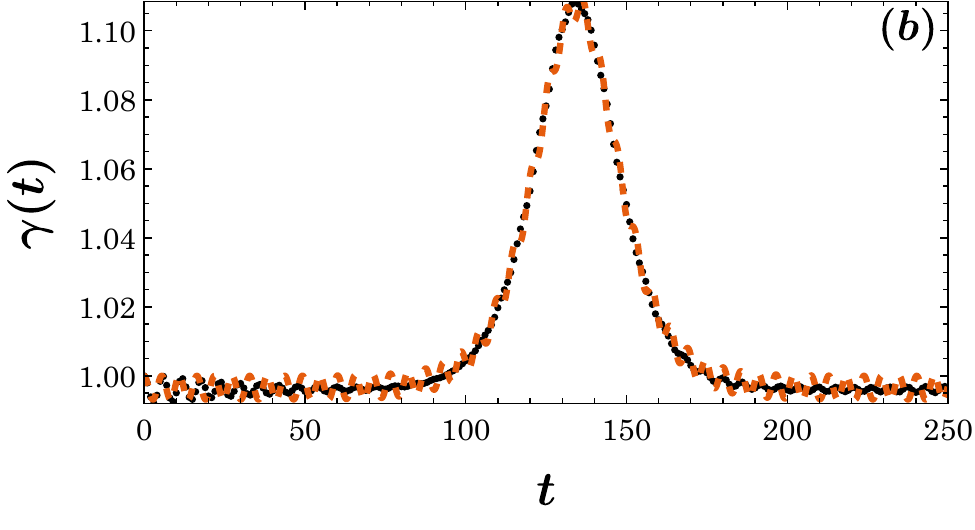}}}
    \quad
    \subfloat{{\includegraphics[height=4.5cm]{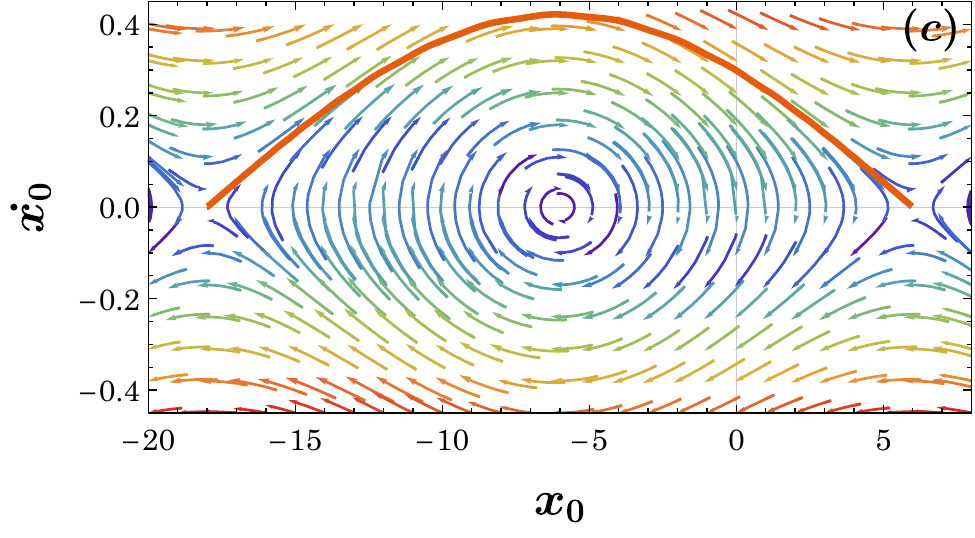}}}
    \caption{Figure (a): A comparison of the kink's position for the solution obtained from the original field model \eqref{sine-gordon} (black line) and the model described by equation \eqref{2dof_ansatz} (orange dashed line) in the periodic potential. Figure (b): The evolution of the variable $\gamma(t)$ obtained from the PDE model (black points) and the effective model (orange dashed line). Figure (c): Phase portrait of the system in the cross-section $\gamma=1$ and $\Dot{\gamma}=0$. The orange line illustrates the trajectory displayed in the top figure. Here, $\varepsilon_1 = 0.1$, $\varepsilon_2 = 0$, $k=\pi/12$, the initial velocity is $v=0.001$, and $x_0=-18$.}
    \label{fig_02}
\end{figure}

\begin{figure}[h!]
    \centering
    \subfloat{{\includegraphics[height=4.5cm]{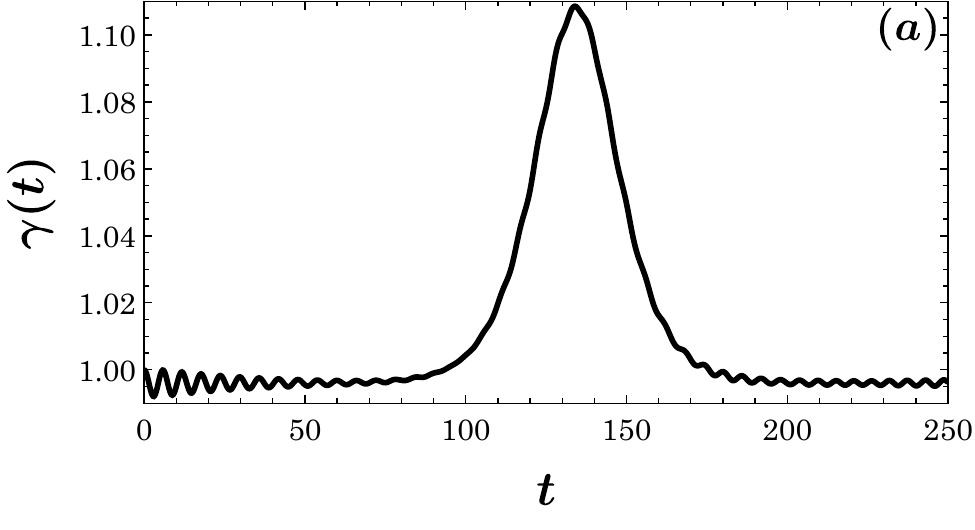}}}
    \quad
    \subfloat{{\includegraphics[height=4.5cm]{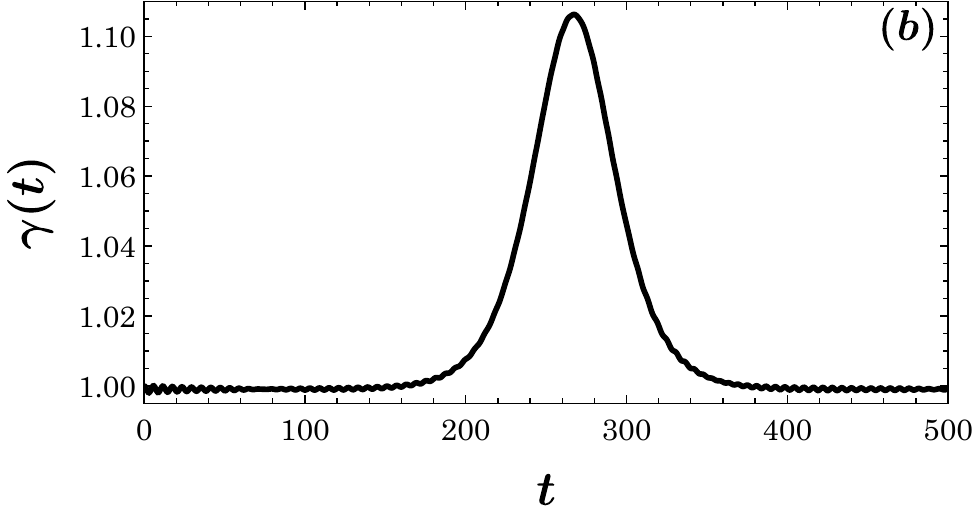}}}
    \caption{{Behavior of the $\gamma$ variable (obtained in the effective model) when the kink moves between maxima of the $g$ function. The parameters in the figure are as follows $\varepsilon_1 = 0.1$, $\varepsilon_2 = 0$ and  $v=0.001$. The function $g$ in figure (a) changes faster and therefore $k=\pi/12$, and $x_0=-18$ while in figure (b) it changes slower so  $k=\pi/24$, and $x_0=-36$.}}
    \label{fig_02v2}
\end{figure}

The next figure illustrates the process of kink transition from one maximum of the function $g$ to the neighboring maximum. In Figure \ref{fig_02} the simulation was performed for the parameters $\varepsilon_1=0.1$, $\varepsilon_2=0$, $k = \pi/12$.  Initially, the kink was placed at one of the maxima (for $x_0=-18$) and assigned a slight velocity in the direction of the neighboring maximum $v=0.001$. Panel (a) of this figure shows the trajectory of the kink. On the right, there is a legend explaining the shape of the function $g$. As one can see, the agreement between the trajectory obtained based on the field model \eqref{sine-gordon} (black line) and the one obtained from the effective model \eqref{2dof_ansatz} (orange dashed line) is highly satisfactory. Panel (b) of this figure shows the evolution of the second variable present in the effective model. The black dots in this diagram were obtained using the field model, while the orange dashed line was obtained based on the effective model. At the beginning, when the kink velocity is minimal, this variable is almost equal to $1$, analogously to the Lorentz factor. 
It is interesting to point out that a very weak oscillation is present
during this interval at the ODE level which does not significantly affect
the quality of the relevant agreement. { The presence of these small oscillations can be explained by the fact that the proposed ansatz (and consequently the initial conditions based on it) approximates the energy minimizing configuration more accurately when the spatially dependent functions appearing in the Lagrangian vary more slowly. This property is illustrated in Figure \ref{fig_02v2}. The figure shows the kink moving between neighboring maxima. Panel (a) presents the result of the effective model in the same situation as depicted in Figure \ref{fig_02} (b). As before, the initial velocity is minimal and equals $v=0.001$. The parameters are respectively $\varepsilon_1=0.1$ and $\varepsilon_2=0$. A similar process is shown in Figure \ref{fig_02v2} (b). The only difference lies in the fact that in panel (b), the function $g(x)$ is more spatially stretched, resulting in a slower spatial variation. Due to the different spatial extension, the maximum in panel (a) is located at $x_0=-18$, while in panel (b) it is located at $x_0=-36$. These positions are also the initial positions of the kink. In both figures, one can observe subtle oscillations emerging at the onset of the evolution. These oscillations arise due to an excess of energy initially present in the configuration defined by the chosen initial conditions. It can be noticed that when the spatial inhomogeneity is more mild the initial oscillation of the kink thickness is much smaller. In this case, the changes in the thickness of the kink, do not have a kinematic character, but are almost completely dynamic in nature (note that the initial velocity is very low).} In both cases
as the velocity increases during movement, the $\gamma(t)$ variable also changes to a value slightly exceeding $1.1$ maximally, before returning to near
unit values. Ignoring dynamic effects, this would correspond to a speed of approximately $0.42$. The correspondence of the two curves is very good here
and, indeed, throughout the trajectory. Panel (c) of Figure \ref{fig_02}, on the other hand, shows the phase portrait of the system in cross-section $(\gamma=1, \Dot{\gamma}=0)$. The solid orange line in this figure represents the trajectory shown in panel (a) of the same figure. 

\FloatBarrier

\subsubsection{The case of $\varepsilon_2 > 0$ and $\omega = 0$}

\begin{figure}[h!]
    \centering
    \subfloat{{\includegraphics[height=4.5cm]{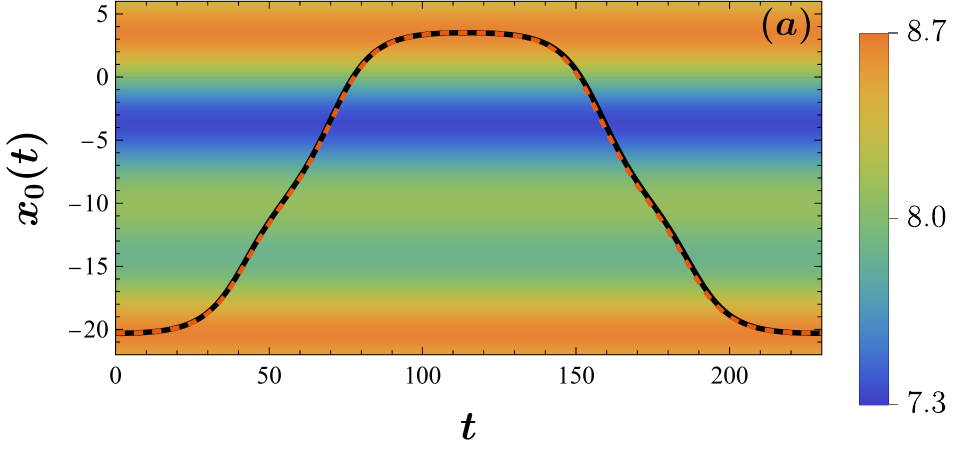}}}
    \quad
    \subfloat{{\includegraphics[height=4.5cm]{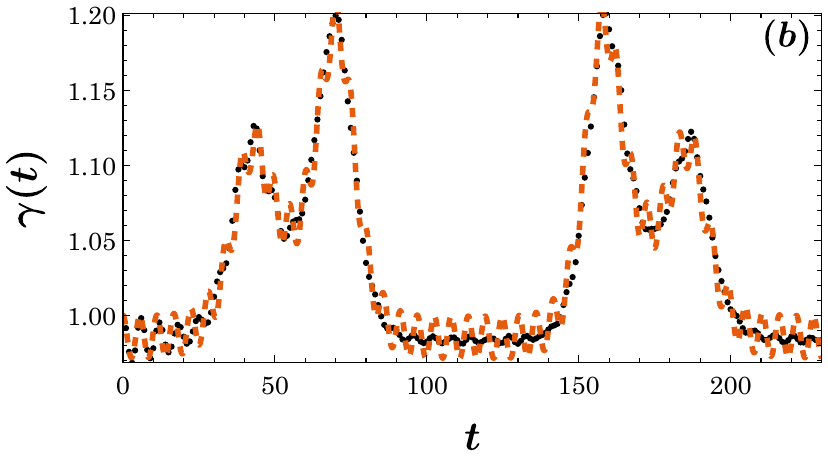}}}
    \quad
    \subfloat{{\includegraphics[height=4.5cm]{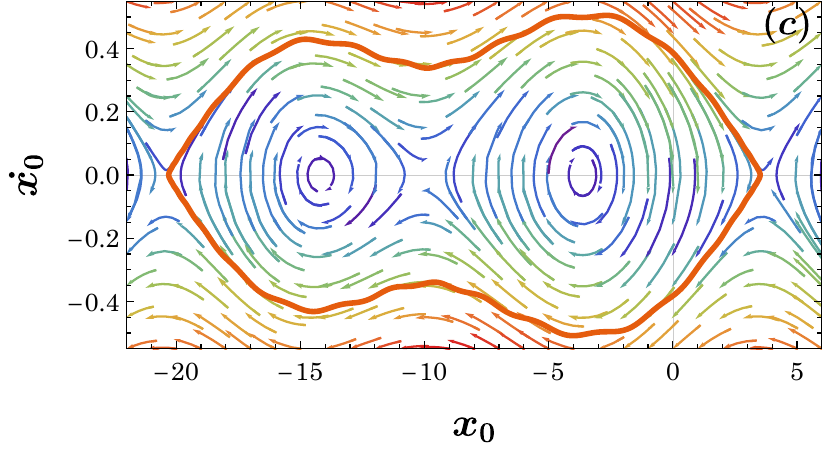}}}
    \caption{Figure (a): A comparison of the kink's position for the solution obtained from the original field model \eqref{sine-gordon} (black line) and the model described by equation \eqref{2dof_ansatz} (orange dashed line) in the periodic potential. Figure (b): The evolution of the variable $\gamma(t)$ obtained from the PDE model (black points) and the effective model (orange dashed line). Figure (c): Phase portrait of the system in the cross-section $\gamma=1$ and $\Dot{\gamma}=0$. The orange line illustrates the trajectory displayed in the top figure. Here, $\varepsilon_1 = 0.1$, $\varepsilon_2 = 0.1$, $k=\pi/6$, the initial velocity is $v=0$, and $x_0=-20.3$.}
    \label{fig_03}
\end{figure}

Both the shape of the effective potential and the dynamics of the system become more complicated when there is a non-trivial contribution from the gradient field in the system. In this case, $\varepsilon_2$ is non-zero, which means that the functions $\cal{F}$ and $g$ have the form $\mathcal{F}(x,t) = 1 + \varepsilon_2\sin\left(\frac{ \pi }{12} x\right)$ and $g(t,x) =1+\varepsilon_1\sin(k x)$. Due to the zeroing of the frequency $\omega$, the system is still an autonomous one. The form of the potential (see the legend located on the right side of the top panel) and the trajectory in the case in consideration are shown in Figure \ref{fig_03}.
The parameter values used in the simulations are $\varepsilon_1=0.1$, $\varepsilon_2=0.1$ and $k=\pi/6$. Panel (a) shows the transition of a kink from one maximum to another maximum of the same height through a maximum of lower height located between them. On the right side of the panel, there is a legend describing the form of the effective  potential \eqref{integrals-fin};
see the details in Appendix A. Strictly speaking, the kink initially rests at the position $x_0=-20.3$ which is close to the potential maximum. From this position, the kink can slide down, overcome the lower (local) maximum it encounters along the way, and finally reach almost the neighboring maximum of the same height.
Panel (a) shows a comparison of the trajectory obtained based on the field model \eqref{sine-gordon} (black line) with the trajectory obtained from the effective model \eqref{2dof_ansatz} (dashed orange line). It can be seen that the effective model accurately reproduces the results of the field equation. Panel (b) shows the time dependence of the variable describing the inverse of the thickness of the kink. The results obtained from the field equation \eqref{sine-gordon} are represented by black dots, while the orange dashed line was obtained based on the effective model \eqref{2dof_ansatz}. The correlation of both lines is remarkably good, although once again the relevant ODE for $\gamma(t)$
features the small amplitude effective width oscillations that are
more pronounced (although still present) at the PDE level. Panel (c) in the figure shows the trajectory (orange line) superimposed on the phase portrait (in the plane $\gamma=1$, $\Dot{\gamma}=0$).

\FloatBarrier

\subsection{Evolution in a non-autonomous system}
\begin{figure}[h!]
    \centering
    \subfloat{{\includegraphics[height=4cm]{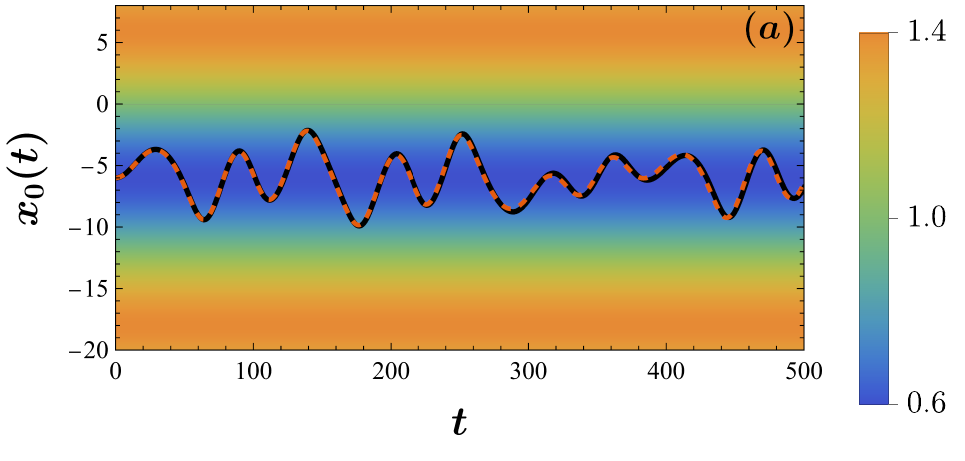}}}
    \quad
    \subfloat{{\includegraphics[height=4cm]{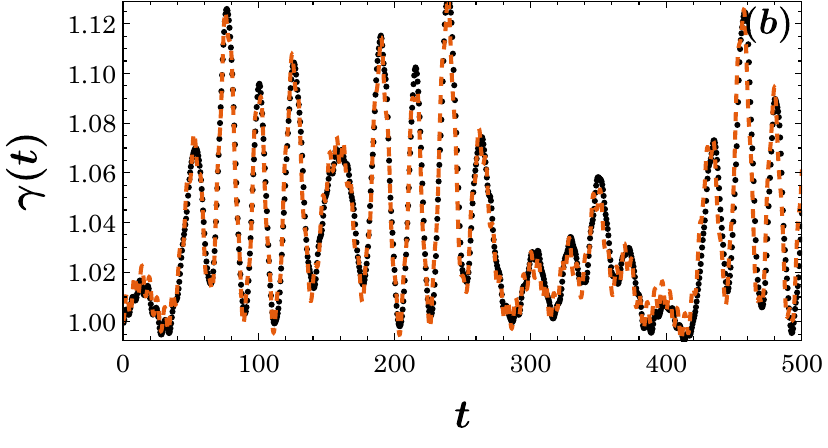}}}
    \caption{Figure (a): A comparison of the kink's position for the solution obtained from the original field model \eqref{sine-gordon} (black line) and the model described by equation \eqref{2dof_ansatz} (orange dashed line). Figure (b): The evolution of the variable $\gamma(t)$ obtained from the PDE model (black points) and the effective model (orange dashed line). Here, $\varepsilon_1 = 0.1$, $\varepsilon_2 = 0.4$, $k=\pi/6$, $\omega=0.05$ the initial velocity is $v=0$, and $x_0=-6$.}
    \label{fig_04}
\end{figure}

The behavior of the system becomes much more interesting when we take into account the time dependence that appears in the $g$ function, i.e., when we assume $\omega > 0$ (using also $\varepsilon_1> 0$). The frequency of the disturbance is assumed to be equal to $\omega=0.05$.
Figure \ref{fig_04} shows the oscillations around the minimum of the function $\mathcal{F}$. In the simulations, the kink initially rests at  $x_0=-6$. As can be seen, due to the time dependence in the $g$ function, the course of these oscillations is highly non-trivial and leads to a far more complex
trajectory of the kink center, and an accompanying highly oscillatory
width variation. In the simulation, the following parameters were assumed: $\varepsilon_1=0.1$  $\varepsilon_2=0.4$ and $k=\pi/6$. This time, the legend refers to the values of the $\mathcal{F}$ function. The black solid line in panel (a) represents the kink position obtained in the field model \eqref{sine-gordon}, while the orange dashed line was obtained in the effective model \eqref{2dof_ansatz}. It should be noted that the similarity of the two models is striking, especially since it concerns very long times (even $t=500$).
The course becomes even more complex in the case of the $\gamma$ variable, the time dependency of which is shown in panel (b). In this panel, the black dots were obtained from the field model, while the dashed orange line represents the result obtained in the effective model. The similarity of the two graphs also in this (far more ``demanding'') case example shows the 
high quality of the effective model as a tool for capturing the
effective kink dynamics in both autonomous and non-autonomous settings.

\begin{figure}[h!]
    \centering
    \subfloat{{\includegraphics[height=4cm]{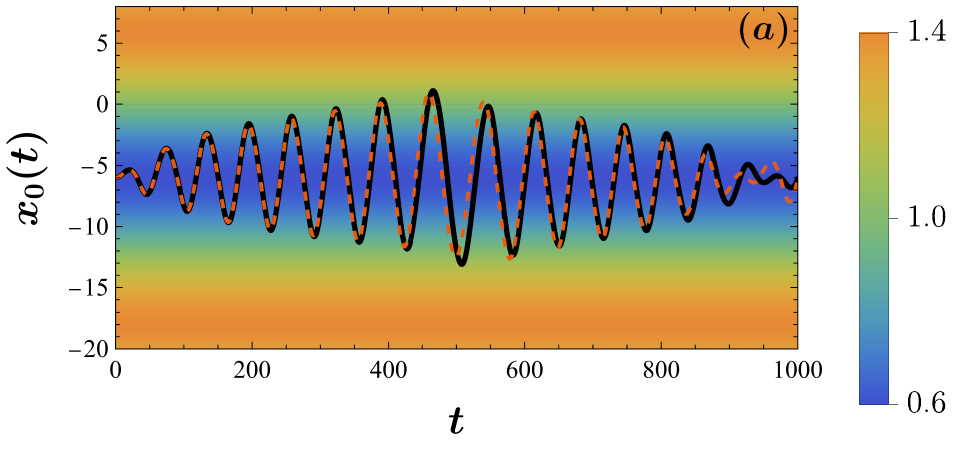}}}
    \quad
    \subfloat{{\includegraphics[height=4cm]{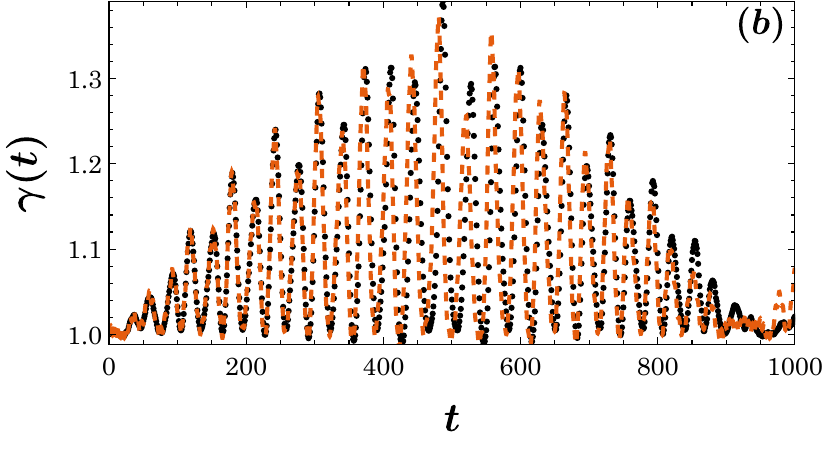}}}
    \caption{Figure (a): A comparison of the kink's position for the solution obtained from the original field model \eqref{sine-gordon} (black line) and the model described by equation \eqref{2dof_ansatz} (orange dashed line). Figure (b): The evolution of the variable $\gamma(t)$ obtained from the PDE model (black points) and the effective model (orange dashed line). Here, $\varepsilon_1 = 0.05$, $\varepsilon_2 = 0.4$, $k=\pi/6$, $\omega=0.05$ the initial velocity is $v=0$, and $x_0=-6$.}
    \label{fig_05}
\end{figure}

 Figure \ref{fig_05} also shows oscillations around the equilibrium position, but this time it resembles a beat. The parameters in the diagram are equal to 
$\varepsilon_1 = 0.05$, $\varepsilon_2 = 0.4$, $k=\pi/6$ and $\omega=0.05$.
This time, the kink also initially rests at $x_0=-6$.
Panel (a) of this figure contains a comparison of the trajectory obtained on the basis of the field model \eqref{sine-gordon} (black line) with the trajectory obtained from the effective model \eqref{2dof_ansatz} (orange dashed line). On the right, there is a legend referring to the values of the $\mathcal{F}$ function. As can be seen, the two graphs are highly consistent, even for times 
reaching many hundreds of units!
Panel (b) of this figure contains a comparison of the time dependencies of the $\gamma$ variable obtained from the field model (black dots) and the effective model (orange dashed line). The remarkable agreement between the reduced
ODE model and the original PDE field theory persists for this diagnostic
too.
 
\begin{figure}[h!]
    \centering
    \subfloat{{\includegraphics[height=4cm]{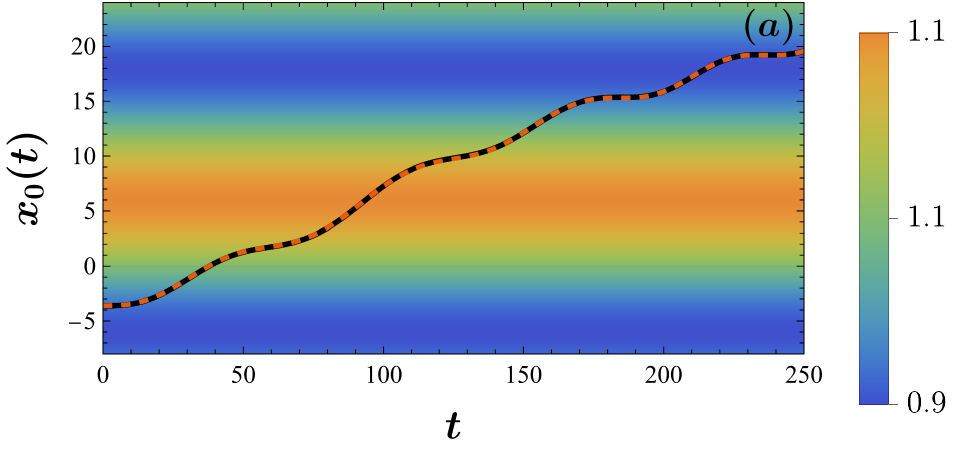}}}
    \quad
    \subfloat{{\includegraphics[height=4cm]{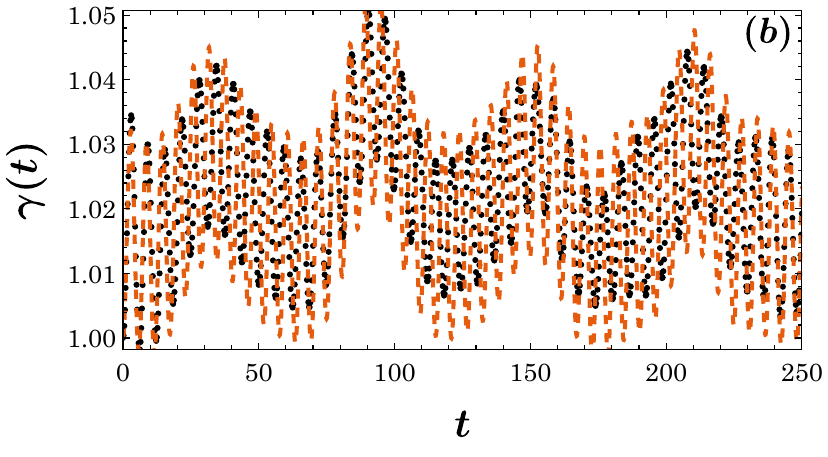}}}
    \caption{Figure (a): A comparison of the kink's position for the solution obtained from the original field model \eqref{sine-gordon} (black line) and the model described by equation \eqref{2dof_ansatz} (orange dashed line). Figure (b): The evolution of the variable $\gamma(t)$ obtained from the PDE model (black points) and the effective model (orange dashed line). Here, $\varepsilon_1 = 0.1$, $\varepsilon_2 = 0.1$, $k=\pi/6$, $\omega=0.05$ the initial velocity is $v=0$, and $x_0=-3.6$.}
    \label{fig_06}
\end{figure}

 Figure \ref{fig_06} shows the process of pushing the kink over the potential barrier in the presence of time-dependent function $g$. The simulations were carried out for the parameters  $\varepsilon_1 = 0.1$, $\varepsilon_2 = 0.1$, $k=\pi/6$ and $\omega=0.05$. Also here, the legend refers to the values of the $\mathcal{F}$ function. 
 {The case presented in this figure is interesting because both the kink movement and the trajectory oscillations are not caused by the presence of 
 a bias current (which is absent here), but are merely a consequence of the dependence of the function $g$ on time. Indeed, it is relevant to recall that the 
 latter has the character of a traveling wave. Importantly, the figure shows that despite the existence of a physical barrier described by the function $\mathcal{F}$, the time dependence of the function $g$ causes the kink to overcome the relevant potential energy barrier.}
Note that the most significant difference between this figure and the two previous ones is the substantially lower value of the parameter $\varepsilon_2$, thanks to which the height of the barrier has been significantly reduced.
Panel (a) of this figure shows a comparison of the kink trajectory obtained from the field equation \eqref{sine-gordon} (black line) with the trajectory obtained from the effective model \eqref{2dof_ansatz} (dashed orange line). The figure shows that they are essentially indistinguishable. Panel (b) of this figure compares the time dependencies of the $\gamma$ variable obtained from the  field equation (black dots) with the values obtained from the effective equation (orange dashed line). Despite the complex multi-frequency nature
of the associated oscillations there is excellent qualitative 
(and reasonable quantitative) agreement between the ODE and PDE models.
 
\FloatBarrier

\section{The System With Dissipation}

\subsection{Effective description of the system with dissipation}

One of the methods  for describing systems in the presence of dissipation is the so-called non-conservative variational method \cite{Galley2013,Kevrekidis2014}. Classically, the variational principle is formulated as a boundary problem (in time) with fixed values of dynamic variables at the beginning and at the end of the evolution. This approach makes it impossible to consider non-symmetric
(i.e., irreversible) processes. In order to allow the formulation of the variational principle to consider such processes, the work of~\cite{Galley2013}
proposed to fix values only at the beginning of the evolution, but not
at the end. In this approach,  the dynamic variables are duplicated which allows the elimination of the boundary condition imposed on the dynamic variables at the final moment. The appropriate construction of the Lagrangian makes it possible to formulate the variational problem in such a way that the variations of the dynamic variables at the beginning of the evolution are determined, while at the final moment the variations of both variables are arbitrary (although ultimately set to be equal to each other in the so-called
``physical limit''). More precisely, the Lagrangian density 
of this non-conservative variational formulation has the form:
\begin{equation}
    \label{nonconservative}
    {\cal L}_N = {\cal L}(\phi_1) - {\cal L}(\phi_2) + {\cal R} .
\end{equation}
The non-conservative Lagrangian density ${\cal L}_N$ contains the combination of the density of the Lagrangians for the first $\phi_1$ and  and second 
field variable $\phi_2$, as well as a term describing the action of 
irreversible (including, e.g., dissipative) processes in the system ${\cal R}$. The analytical form of both Lagrangian densities is like the one shown in the equation \eqref{L}.
It is worth emphasizing that
the auxiliary variables $\phi_1$ and $\phi_2$ after the extremization procedure are enforced to
coincide with the physical variable $\phi$; this is the physical limit.
The process of identifying the two trajectories and that of extremization of the action do not commute and hence the procedure
enables the incorporation of non-conservative forces in a Lagrangian formulation.

In this approach one can separate the conservative part of the equation of motion from the non-conservative component
\begin{equation}
    \label{eq-nonconservative}
    \partial_{\mu} \left( \frac{\partial {\cal L}}{\partial (\partial_{\mu} \phi)}\right) - \frac{\partial {\cal L}}{\partial \phi} = \left[\frac{\partial {\cal R}}{\partial \phi_{-}} - \partial_{\mu} \left( \frac{\partial {\cal R}}{\partial (\partial_{\mu} \phi_{-})}\right)\right]_{PL},
\end{equation}
where the index $\mu$ enumerates the space-time variables $x^{\mu}=(x^0,x^1)=(t,x)$ and we assumed the standard summation convention.
Moreover we used variables, $\phi_{-}$ as well as $\phi_{+}$ which are related to the original variables $\phi_1$ and $\phi_2$ as follows: $\phi_1=\phi_{+}+ \frac{1}{2}\phi_{-}$ and $\phi_2=\phi_{+}- \frac{1}{2}\phi_{-}$
(or conversely $\phi_+=(\phi_1+\phi_2)/2$ and $\phi_-=\phi_1-\phi_2$).
Additionally, the inscription PL denotes the physical limit, in which $\phi_{+}$ becomes a physical variable $\phi_{+}=\phi$ and $\phi_{-}=0$. 
For the Lagrangian density \eqref{L}, considered in this article the equation \eqref{eq-nonconservative} can be converted to the form
\begin{equation}
\label{eq-nonconservative2}
    \partial_t^2 \phi - \partial_x (\mathcal{F}(x)\partial_x \phi) + \sin \phi = \left[\frac{\partial {\cal R}} {\partial \phi_{-}} - \partial_{\mu} \left( \frac{\partial {\cal R}}{\partial (\partial_{\mu} \phi_{-})}\right)\right]_{PL} .
\end{equation}
Taking the non-conservative part of the Lagrangian in the form 
\begin{equation}
   {\cal R} = -\Gamma \phi_{-} - \eta \phi_{-} \partial_t \phi_{+} ,
\end{equation}
 we can reproduce the equation \eqref{sine-gordon+}. We obtain the effective description of the system by inserting the Ansatz \eqref{xi} and \eqref{phi} into the expression for ${\cal R}$ and upon integrating over the spatial variable i.e. $R_{eff} = \int_{-\infty}^{+\infty} dx {\cal R}$. The effective equations of motion can then be written as follows
\begin{equation}
    \label{eff-eq1}
    \frac{d}{d t }\left(\frac{\partial L_{eff}}{\partial \Dot{x}_0} \right) - \frac{\partial L_{eff}}{\partial {x}_0} = \left[ \frac{\partial R_{eff}}{\partial {x}_{-}} -
    \frac{d}{d t }\left(\frac{\partial R_{eff}}{\partial \Dot{x}_{-}} \right) \right]_{PL} ,
\end{equation}
\begin{equation}
    \label{eff-eq2}
    \frac{d}{d t }\left(\frac{\partial L_{eff}}{\partial \Dot{\gamma}} \right) - \frac{\partial L_{eff}}{\partial {\gamma}} = \left[ \frac{\partial R_{eff}}{\partial {\gamma}_{-}} -
    \frac{d}{d t }\left(\frac{\partial R_{eff}}{\partial \Dot{\gamma}_{-}} \right) \right]_{PL} ,
\end{equation}
where $L_{eff}$ was obtained in the previous section \eqref{Leff}.
This time in the physical limit (for the reduced variables $x_0$
and $\gamma$) we have $x_{-}=0$, $\gamma_{-}=0$, $x_{+}=x_0$ and $\gamma_{+}=\gamma$.
Moreover, the relationships for effective variables are analogous to those for field variables i.e.,
$x_1=x_+ + \frac{1}{2} x_-$, $x_2=x_+ - \frac{1}{2} x_-$ and $\gamma_1=\gamma_+ + \frac{1}{2} \gamma_-$, $\gamma_2=\gamma_+ - \frac{1}{2} \gamma_-$.
The system of ordinary differential equations for the effective variables takes the final form
\begin{equation}
\begin{gathered}
\label{2dof_ansatz-dyss}
    M\Ddot{x}_0-\kappa\Ddot{\gamma}+\frac{1}{2}(\partial_{x_0}M)\Dot{x}_0^2-\frac{1}{2}(\partial_{x_0}m)\Dot{\gamma}^2-(\partial_{\gamma}\kappa)\Dot{\gamma}^2+(\partial_{\gamma}M)\Dot{\gamma}\Dot{x}_0+ \\
    (\partial_t M) \Dot{x}_0 - (\partial_t \kappa) \Dot{\gamma} + (\partial_{\gamma} \alpha) \Dot{\gamma} + (\partial_{x_0} \beta) \Dot{\gamma} + \partial_t \alpha +   \partial_{x_0}V=\\
 2 \pi \Gamma - \eta \frac{\pi^2}{6 \gamma} \sqrt{\frac{{\cal F}(x_0)}{g(t,x_0)}} \left( \frac{\partial_{x_0} {\cal F}(x_0)}{{\cal F}(x_0)} - \frac{\partial_{x_0} g(t,x_0)}{g(t,x_0)}\right)^2 \Dot{x}_0 - 8 \eta \gamma \sqrt{\frac{g(t,x_0)}{{\cal F}(x_0)}} \Dot{x}_0 + \\ \eta \frac{\pi^2}{3 \gamma^2} \sqrt{\frac{{\cal F}(x_0)}{g(t,x_0)}} \left( \frac{\partial_{x_0} {\cal F}(x_0)}{{\cal F}(x_0)} - \frac{\partial_{x_0} g(t,x_0)}{g(t,x_0)}\right) \Dot{\gamma} -
\eta \frac{\pi^2}{6} \frac{f(t,x_0)}{\gamma g(t,x_0)}\sqrt{\frac{{\cal F}(x_0)}{g(t,x_0)}} \left( \frac{\partial_{x_0} {\cal F}(x_0)}{{\cal F}(x_0)} - \frac{\partial_{x_0} g(t,x_0)}{g(t,x_0)}\right)
    ,
    \\
    m\Ddot{\gamma}-\kappa\Ddot{x}_0 +\frac{1}{2}(\partial_{\gamma}m)\Dot{\gamma}^2-\frac{1}{2}(\partial_{\gamma}M)\Dot{x}_0^2-(\partial_{x_0}\kappa)\Dot{x}_0^2+(\partial_{x_0}m)\Dot{x}_0\Dot{\gamma}+\\
(\partial_t m) \Dot{\gamma} - (\partial_t \kappa) \Dot{x}_0 - (\partial_{x_0} \beta) \Dot{x}_0 - (\partial_{\gamma} \alpha) \Dot{x_0} - \partial_t \beta +   \partial_{\gamma}V=\\
\eta \frac{2 \pi^2}{3 \gamma^2} \sqrt{\frac{{\cal F}(x_0)}{g(t,x_0)}} \left[ \left( \frac{\partial_{x_0} {\cal F}(x_0)}{{\cal F}(x_0)} - \frac{\partial_{x_0} g(t,x_0)}{g(t,x_0)}\right) \Dot{x}_0  - \frac{\Dot{\gamma}}{\gamma} + \frac{f(t,x_0)}{2 g(t,x_0)}\right] . 
\end{gathered}
\end{equation}
All  parameters present in the above equations are defined in the Appendix.

\subsection{Numerical results for the system with dissipation}
\begin{figure}[h!]
    \centering
    \subfloat{{\includegraphics[height=4cm]{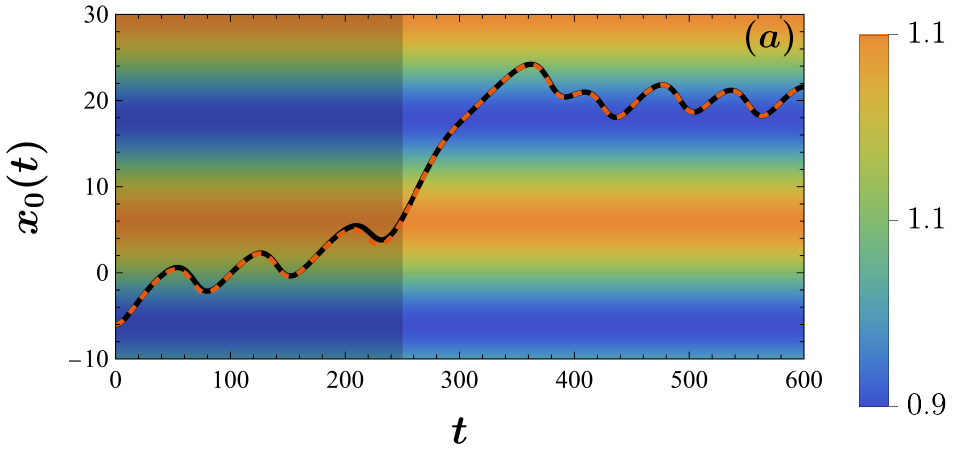}}}
    \quad
    \subfloat{{\includegraphics[height=4cm]{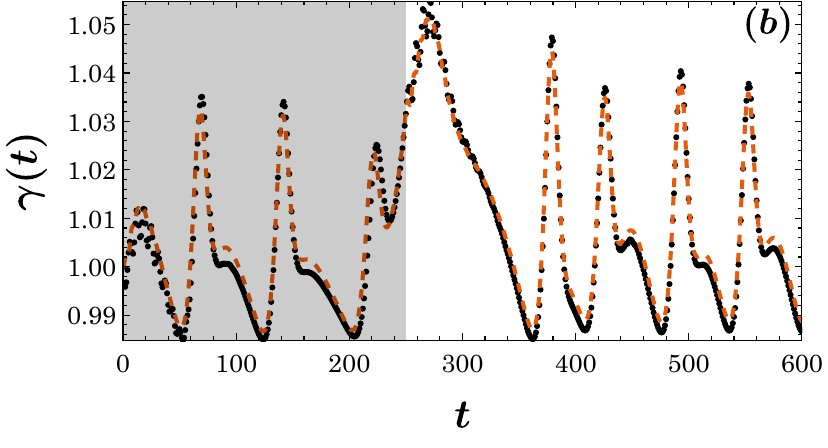}}}
    \caption{Figure (a): A comparison of the kink's position for the solution obtained from the original field model \eqref{sine-gordon+} (black line) and the model described by equation \eqref{2dof_ansatz-dyss} (red dashed line). Figure (b): The evolution of the variable $\gamma(t)$ obtained from the PDE model (black points) and the effective model (orange dashed line). Here, $\varepsilon_1 = 0.1$, $\varepsilon_2 = 0.1$, $k=\pi/6$, $\omega=0.1$ the initial velocity is $v=0$, $x_0=-6$, $\eta =0.1$, and $\Gamma=0.01$ for $0 \leq t \leq 250$ and $\Gamma=0$ otherwise.}
    \label{fig_07}
\end{figure}

An example of the behavior of a kink in a system in which both dissipation and external forcing in the form of a bias current occur is shown in Figure \ref{fig_07}. Panel (a) of this figure  compares the kink behavior obtained based on the field equation \eqref{sine-gordon+}  and the effective model \eqref{2dof_ansatz-dyss}.
The legend on the right refers to the values of the $\mathcal{F}$ function. The simulation is carried out using the parameters $\varepsilon_1=0.1$, $\varepsilon_2=0.1$, $k=\pi/6$ and $\omega=0.1$. The dissipation coefficient is equal to $\eta=0.1$.
In addition, from the beginning of the simulation until time $t=250$, there is a forcing in the system in the form of a "bias current".
Once again, the kink trajectory obtained based on the field equation has the form of a black line, while the trajectory obtained from the effective equation is represented by an orange dashed line.
Initially, the kink is at rest ($v=0$) at the point $x_0=-6$. The time dependence of function g causes oscillations in the position of kink. These oscillations are too small to shift the kink to the neighboring minimum of function $\mathcal{F}$.
This movement occurs due to an external force in the form of a bias current. The duration of the external forcing is indicated in the figure by a gray field. 
Note that after shifting to the vicinity of the second minimum, despite the existence of dissipation in the system, the amplitude of the oscillations does not decrease over time. This behavior has its origin in the time dependence of the $g$ function. The panel (a) of the figure shows the remarkable agreement of the two trajectories, even for times up to $t=600$.
Panel (b) of this figure shows the behavior of the $\gamma$ variable during the transition from one minimum to another. The black dots represent the prediction from the field model, while the orange dashed line is the result of the effective model. It can be seen that the two curves are very well matched,
despite the complex time evolution involving multiple frequencies. The apparent oscillations correspond to the variations in the g function, while the central peak in the graph is related to the kink sliding down from the barrier separating the two minima.

\FloatBarrier

\section{Conclusions and Future Challenges}
In our current work, we have conducted explored the role of various spatial inhomogeneities on the dynamics of the kink in a perturbed sine-Gordon model. A new element in the research is the assumption that the system is not autonomous. The appearance of an explicit dependence of the coefficients in the equation on time introduces interesting possibilities, such as the ability to manipulate/guide the kink across wells of a periodic potential, or to introduce beating
phenomena in both the position and the width of the associated coherent 
structure. Understanding these behaviors is possible within the proposed effective model. The model is the result of leveraging a somewhat unusual ansatz, in that it explicitly factors 
in the role of the (spatio-temporal) inhomogeneities 
acting on the PDE problem. As is common in such settings~\cite{CKW14}, our Ansatz is based on two variables. The first variable describes the position of the kink understood as the location of its center, i.e., the place where the scalar field takes the value of $\pi$. The second variable describes the inverse of the kink thickness. In some situations described in the paper, this variable takes values close to the value of the Lorentz factor. 

The  description proposed herein yields satisfactory results  both in the case where dissipation in the system is absent and in the case where dissipation
plays a significant role. In the latter case, we used an approach based on the non-conservative variational approximation introduced in \cite{Galley2013}
(and further discussed in~\cite{galley2014principlestationarynonconservativeaction}). It is important that the suggested method of constructing an effective model leads to
a remarkable agreement of the obtained ODE trajectories in comparison with the trajectories determined on the basis of the original field model. This agreement reaches hundreds or even thousands of time units. This is particularly striking when there is an explicit dependence of the model parameters on time, i.e.,
non-autonomous dynamics, and the trajectory is extremely complicated. The disturbance of parameters in question takes the form of a wave running through the system. It turns out that this type of temporal change in the value of the parameters, even without an external forcing, can 
lead to the motion of a kink through the system. This behavior is particularly intriguing when it induces a shift between different equilibrium points of a system in which there is no temporary disturbance. Naturally, a time-dependent external drive can have similar effects, which has also been shown in simulations. Indeed, this forms the basis for the guidance and
manipulation of solitary waves which is a topic of considerable
interest in discrete~\cite{hector}, continuum yet heterogeneous~\cite{theocharis} dispersive wave systems, and also in ones incorporating gain and
loss features, both theoretically~\cite{Rossi2024}, but, importantly, also
experimentally~\cite{Jang2015}.

Naturally, the work paves the way for further research into the described issues. Potential directions include the detailed 
study of parametric resonance, as it can be induced from these
non-autonomous perturbations to the equations of motion.
Accordingly, the availability of this highly efficient reduced description
that systematically captures the dynamics of the problem opens new
directions for the consideration, control, guidance and manipulation
of kink dynamics at that highly reduced level, before applying relevant
ideas to the full field theoretic model.
As such, the issues of existence, stability, chaotic behavior and guided
motion in the two-degree-of-freedom models, as well as  the modification of field equations with terms that describe, for example, the surface impedance of the Josephson junction, become relevant. More generally, this
also paves the way for generalizing similar considerations
to higher dimensions which are of interest in their own right, especially
in light of the kink's robustness in the latter settings. Such directions
are currently under consideration and will be reported in future publications.

\section{Appendix}
The parameters present in the effective Lagrangian \eqref{Leff} are defined by the integrals below
\begin{equation}
\begin{gathered}
\label{integrals}
    M = 4\int_{-\infty}^{+\infty}dx\sech^2(\xi) \, W^2(\xi),\\
    m = \frac{4}{\gamma^2}\int_{-\infty}^{+\infty}dx\sech^2(\xi) \, \xi^2,\\
    \kappa = \frac{4}{\gamma} \int_{-\infty}^{+\infty}dx\sech^2(\xi) W(\xi) \, \xi ,\\
\alpha = 2\int_{-\infty}^{+\infty}dx\sech^2(\xi) \, \frac{f(t,x_0)}{g(t,x_0)} \, W(\xi) \xi
    , \\
\beta = \frac{2}{\gamma} \int_{-\infty}^{+\infty}dx\sech^2(\xi) \, \frac{f(t,x_0)}{g(t,x_0)} \, \xi^2
    ,\\
    V = 2\int_{-\infty}^{+\infty}dx\sech^2(\xi) \, \left( g(t,x) + g(t,x_0) \, \frac{\mathcal{F}(x)}{\mathcal{F}(x_0)}\gamma^2 - \frac{f^2(t,x_0)}{4 g^2(t,x_0)} \, \xi^2 \right) .
\end{gathered}
\end{equation}
Auxiliary functions $W$ and $f$ appearing in expressions for effective Lagrangian coefficients are as follows
\begin{equation} \label{W}
    W(\xi) = \frac{1}{2} \, \left( \frac{\partial_{x_0} {\cal F}(x_0)}{ {\cal F}(x_0)} - \frac{\partial_{x_0} { g}(t,x_0)}{ {g}(t,x_0)} \right) \, \xi + \sqrt{\frac{g(t,x_0)}{{\cal F}(x_0)}} \, \gamma , \,\,\,\,\, f(t,x_0) = \omega \varepsilon_1 \cos{(k x_0 - \omega t)} .
\end{equation}
All coefficients are functions of the dynamic variables $x_0(t)$ and $\gamma(t)$. Moreover, they depend explicitly on time.
Most of the above integrals can be calculated by obtaining the explicit dependence of the coefficients on the functions ${\cal F}(x_0)$ and $g(t,x_0)$ present in the lagrangian density and on their derivatives
\begin{equation}
\begin{gathered}
\label{integrals-fin}
    M = \frac{\pi^2}{6 \gamma} \sqrt{\frac{{\cal F}(x_0)}{g(t,x_0)}} \left( \frac{\partial_{x_0}{\cal F}(x_0)}{{\cal F}(x_0)} - \frac{\partial_{x_0} g(t,x_0)}{g(t,x_0)}\right)^2 + 8 \gamma \sqrt{\frac{g(t,x_0)}{{\cal F}(x_0)}},\\
    m = \frac{2 \pi^2}{3 \gamma^3} \sqrt{\frac{{\cal F}(x_0)}{g(t,x_0)}},\\
    \kappa = \frac{\pi^2}{3 \gamma^2} \sqrt{\frac{{\cal F}(x_0)}{g(t,x_0)}} \, \left( \frac{\partial_{x_0}{\cal F}(x_0)}{{\cal F}(x_0)} - \frac{\partial_{x_0} g(t,x_0)}{g(t,x_0)}\right) ,\\
\alpha = \frac{\pi^2}{6 \gamma} \, \sqrt{\frac{{\cal F}(x_0)}{g(t,x_0)}} \left( \frac{\partial_{x_0}{\cal F}(x_0)}{{\cal F}(x_0)} - \frac{\partial_{x_0} g(t,x_0)}{g(t,x_0)}\right) \frac{f(t,x_0)}{g(t,x_0)}  , \\
\beta = \frac{\pi^2}{3 \gamma^2} \, \sqrt{\frac{{\cal F}(x_0)}{g(t,x_0)}} \frac{f(t,x_0)}{g(t,x_0)}   ,\\
    V = 2 \int_{-\infty}^{+\infty} dx \sech^2 \xi \, g(t,x) +\frac{2 \gamma^2 g(t,x_0)}{{\cal F}(x_0)} \int_{-\infty}^{+\infty} dx \sech^2 \xi \, {\cal F}(x) - \frac{\pi^2 f^2(t,x_0)}{12 \gamma g^2(t,x_0)} \sqrt{\frac{{\cal F}(x_0)}{g(t,x_0)}}  .
\end{gathered}
\end{equation}
Note that only the potential is partially determined by integrals containing functions ${\cal F}(x)$ and $g(t,x)$.

\section{Acknowledgments}
This material is based upon work supported by the U.S. National Science Foundation under the awards PHY-2110030, DMS-2204702 and PHY-2408988 (PGK). Research project supported by program "Excellence Initiative - Research University" for the AGH University of Krakow (JG).

\FloatBarrier
\printbibliography

\end{document}